\documentclass{iau} 
\usepackage{graphicx}

\title[Pulsars at millimetre wavelengths] 
{Pulsar observations at\\millimetre wavelengths}

\author[Pablo Torne]   
{Pablo Torne$^{1,2}$}

\affiliation{
$^1$Instituto de Radioastronom\'ia Milim\'etrica, IRAM \\ Avenida Divina Pastora 7, Local 20, 18012, Granada, Spain \\ email: {\tt torne@iram.es} \\[\affilskip]
$^2$Max-Planck-Institut f\"ur Radioastronomie, MPIfR \\ Auf dem H\"ugel 69, 53121 Bonn, Germany \\
}

\pubyear{2017}
\volume{337}  
\jname{Pulsar Astrophysics -­ The Next 50 Years}
\editors{P. Weltevrede, B.B.P. Perera, L. Levin Preston \& S. Sanidas, eds.}

\begin{document}

\maketitle

\begin{abstract}
Detecting and studying pulsars above a few GHz in the radio band is challenging due to
the typical faintness of pulsar radio emission, their steep spectra, and the lack of observatories
with sufficient sensitivity operating at high frequency ranges. 
Despite the difficulty, the observations of pulsars at high radio frequencies are valuable because they can help us 
to understand the radio emission process, complete a census of the Galactic pulsar population, and possibly discover
the elusive population in the Galactic Centre, where low-frequency observations have problems due to the strong scattering.
During the decades of the 1990s and 2000s, the availability of sensitive instrumentation allowed for the detection
of a small sample of pulsars above 10$\,$GHz, and for the first time in the millimetre band.
Recently,
new attempts between 3 and 1$\,$mm ($\approx$86$-$300$\,$GHz) have resulted in the detections of a pulsar
and a magnetar up to the highest radio frequencies to date, reaching 291$\,$GHz (1.03$\,$mm).
The efforts continue, and the advent of new or upgraded millimetre facilities like the IRAM 30-m, NOEMA, 
the LMT, and ALMA, warrants a new era of high-sensitivity millimetre pulsar astronomy in the upcoming years.

\end{abstract}

\firstsection 
\section{Why observe pulsars at millimetre wavelengths?} 

\subsection{Pulsar radiation and the radio emission process}

Most of what we know about pulsars is derived from observations up to $\sim$10$\,$GHz, 
with the vast majority of the available data obtained below 2$\,$GHz (\cite[Manchester et al. 2005]{Man95}).
At radio frequencies, pulsar spectra are generally steep, with spectral indices typically between 0 and $-$3.5, and mean value $\left<\alpha\right>=-$1.8$\pm$0.2 (\cite[Maron et al. 2000]{Mar00}).
The radiation can have a high degree of linear polarization, reaching in some cases 100\%, sometimes also with a circular component.
The brightness temperatures of the pulsations are well above the thermal limit ($T_{\mathrm{b}}\approx10^{12}\,$K),
with single pulses as narrow as a few nano seconds and reaching up to $T_{\mathrm{b}} \sim10^{41}\,$K (\cite[Hankins \& Eilek 2007]{Han07}).
From the observed brightness temperatures we know that the radio emission mechanism must be coherent.
However, after 50 years of studies of these objects, the exact process leading to coherence is still one of the main open questions in pulsar astronomy.

High-frequency observations, reaching the millimetre and even shorter wavelengths, can help us to identify the radio emission process.
A good sample of pulsars with a dense spectral coverage would provide observational input 
that any valid emission theory must be able to reproduce, becoming a test for the different available models.
Unfortunately, not all the proposed models offer predictions verifiable through observations
(for a review of pulsar emission models, see e.g. \cite[Melrose \& Yuen 2016]{melyun06}).
An example of a model that may be tested is the antenna mechanism produced by bunched particles.
It predicts the loss of efficiency of the coherent process, which translates into a decrease of the intensity, as the frequency increases.
At sufficiently high frequencies, the expected underlying incoherent emission would become dominant, an effect that may be observed
as a turn-up in the spectrum (\cite[Michel 1978]{mich78}).
The loss of coherence should also imply an accompanying depolarization of the radiation.
The transition frequency between the coherent and incoherent components is unknown, but identifying it would provide the important coherence length of the radiative process,
and would support the validity of this model.

An additional relevant model is the so-called radius-to-frequency mapping (\cite[Cordes 1978]{cor78}), 
which proposes a relationship between the wavelength and the emission height in the pulsar magnetosphere.
If the prediction is correct, we may be able to use high-frequency observations to study the magnetospheric 
region close to the neutron star surface, where the higher plasma
densities or a deviation from the assumed dipolar configuration of the magnetic field may imprint observable effects in the radio emission.




\vspace{2mm}

\subsection{Finding new pulsars}

Although increasing our comprehension of the pulsar emission mechanism is a quest where high radio frequencies play an important role,
it is not their only application. We can also use them to try to find new pulsars.
Frequencies around $\sim$1$-$2$\,$GHz are generally the best compromise between pulsar luminosities and interstellar medium (ISM) detrimental effects,
but the regions of the Galaxy behind very dense and turbulent gas may be difficult to survey at these frequencies.
The main cause is the scattering of the radio waves that broadens the pulses decreasing their detectability. 
In extreme cases, the pulsations can be completely washed out at low frequencies, hindering the detections.
In these situations, increasing the observing frequency may be the only way to overcome the scattering.

One particular region where high-frequency surveys can be very useful is the centre of the Milky Way.
Pulsars found in the Galactic Centre can help us to understand its enigmatic star formation history using their characteristics ages, 
map the gravitational potential using the pulsars as accelerometers, measure the gas properties and distribution and estimate the magnetic field through Faraday rotation effects. 
In addition, it is remarkable that one single pulsar in a suitable orbit around the supermassive black hole Sgr$\,$A* would suffice to enable unprecedented
black hole physics experiments and tests of General Relativity and alternative Gravity theories (\cite[Liu et al. 2012]{liu12}, \cite[Psaltis et al. 2016]{psal16}).
A pulsar orbiting Sgr$\,$A* is furthermore a desired complement to the ongoing black hole studies with IR-stars and the event horizon imaging (see e.g. \cite[Goddi et al. 2017]{goddi17}).
Although a large population of pulsars is expected in the Galactic Centre (e.g. \cite[Wharton et al. 2012]{whar12}),
to date only six pulsars have been found in the inner 80 parsecs (\cite[Johnston et al. 2006]{joh16}, \cite[Deneva et al. 2009]{den09}, \cite[Eatough et al. 2013b]{eat13b}).
In the past, the predicted extreme scattering in this particular direction (\cite[Cordes \& Lazio 2002]{corlaz02}) was accepted as the cause for the paucity of detections.
However, the discovery in 2013 of the radio magnetar SGR~J1745$-$2900, 
the closest pulsar to the centre of the Galaxy (\cite[Rea et al. 2013]{rea13}), showed that the scattering was significantly weaker than predicted (\cite[Spitler et al. 2014]{spi14}).
This latter finding in combination with the non-detections of pulsars from the previous many surveys (even using high frequencies, e.g. \cite[Eatough et al. 2013a]{eat13a})
have generated a sense of mystery around the Galactic Centre pulsar population.
In any case, just the fact that the only pulsar close to Sgr$\,$A* is a radio magnetar (one of the rarest types of pulsars),
and that it belongs to the 3\% most luminous pulsars known, strongly suggests that there should be a large still-undetected population in the region.
Upcoming facilities like the Square Kilometre Array, or high-sensitivity surveys at millimetre wavelengths,
should finally be able to uncover the elusive pulsars in the inner few parsecs of the Galaxy.

Finally, the use of (sub)millimetre pulsar surveys can be a suitable method to detect radio emission from magnetars. Being the only pulsars showing inverted radio spectra,
they might be more easily detected at high frequencies than at low frequencies.

\section{What have we learnt so far?}

The completion of the 100-m telescope of the MPIfR at Effelsberg made a big step forward in the observations of pulsars at high frequencies.
During the 1990s, it allowed for detections at unprecedented frequencies of up to 43$\,$GHz (7$\,$mm, for a review, see \cite[Wielebinski et al. 2000]{wie00}).
In a series of papers focused on observations between 30$-$40$\,$GHz, it was shown that the emission could show intrinsic variability (\cite[Kramer et al. 1997]{kra97}),
the flux density of some pulsars was higher than expected from an extrapolation of low-frequency data, suggesting a possible spectral turn-up (\cite[Kramer et al. 1996]{kra96}),
and that the emission can show depolarization at high frequencies (\cite[Xilouris et al. 1996]{xil96}).
Those first observations of pulsars in the millimetre band align surprisingly well with a coherence breakdown, 
as predicted by the bunched particles emission model (Michel 1982).
The record detection at that time was achieved for PSR~B0355+54 with the 30-m telescope of IRAM, in Spain, at 87 GHz (3.4$\,$mm).
Nonetheless, this detection did not show a conclusive turn-up in the spectrum even at those frequencies (\cite[Morris et al. 1997]{morr97}).
This added uncertainty to the previous results about the possible turn-up, and suggests that the transition frequency between
emission components may differ for different pulsars.

In the decade of the 2000s, emission from the highest-magnetic-field neutron stars, or magnetars, was first observed in the radio band (\cite[Camilo et al. 2006]{cam06}).
It took only two years to confirm that the emission from these objects was peculiar.
Some of the emission characteristics were similar to that of the normal pulsar population, like the high brightness temperature of the radio pulses,
but others were very different: radio emission activated after a high-energy outburst, high-energy luminosities too high to be explained solely by spindown,
variable pulse profiles, polarization levels roughly constant with frequency, or a tendency to have a flat or inverted spectrum.
Notably, the latter property offers a unique opportunity to study the properties of pulsar emission in the millimetre band and above.
In fact, this characteristic led to the detection of pulsations from XTE~J1810$-$197 up to 144$\,$GHz (2.08$\,$mm, \cite[Camilo et al. 2007]{cam07}),
setting the record-radio-frequency detection of pulsars for almost a decade.

In recent years, triggered by the discovery of SGR~J1745$-$2900 and the availability of new, large-bandwidth receivers at the IRAM 30-m telescope,
new observational campaigns at millimetre wavelengths resulted in the detection of SGR~J1745$-$2900 up to 291$\,$GHz (1.03$\,$mm),
including single pulses up to 154 GHz (1.95$\,$mm), and measuring a high degree of linear polarization up to 138$\,$GHz (2.17$\,$mm, Torne et al. 2015, 2017). 
PSR~B0355+54 was detected up to 138$\,$GHz (Torne 2017, Torne et al. \emph{in prep.}).
In addition, the first reported imaging observation of the Vela pulsar with ALMA shows detections up to 343$\,$GHz (0.87$\,$mm, Mignani et al. 2017).
However, if the emission is pulsed or not yet requires confirmation.

The recent results complete those reported during the previous decades, and show that pulsars can emit well within the short millimetre band. 
Furthermore, the measured brightness temperatures show that the emission is still coherent up to at least 154 GHz (Torne et al. 2015), and very possibly 343$\,$GHz (Mignani et al. 2017).
A remarkable property seen in the new data for PSR~B0355+54 and SGR~J1745$-$2900 is a strong intrinsic variability. For magnetars it is common to observe variability, 
but for the normal pulsar, this result together with the those of \cite[Kramer et al. 1996]{kra96}, supports the idea that variability could also be a common 
characteristic of all pulsars at very high frequencies.

\section{Summary}

The observations of pulsars at millimetre wavelengths are challenging, but provide unique insights into the pulsar emission properties,
offer a way to probe dense ISM and find new pulsars and magnetars, and are a potential tool for precision black hole physics in the
case that scattering prevents the detection of pulsars orbiting Sgr$\,$A* at low frequencies. 
For this reason, we should continue the efforts to enable sensitive pulsar observations up to the highest possible frequencies.
At the moment, at millimetre wavelengths the observations are concentrated at the IRAM 30-m radio telescope.
In the future, larger facilities like NOEMA or the LMT have the potential to extend the sample of detectable pulsars between 3 and 0.8$\,$mm.
Finally, a key facility will be ALMA, not only because it is the most sensitive (sub)millimetre telescope, 
but also because it is located in the southern hemisphere. Around 70\% of all known pulsars have declination lower than zero degrees, 
and many of them are out of the visibility of the northern facilities. 
Therefore, the future observations with ALMA will offer the best chance to expand our comprehension of the pulsar emission physics,
and have the potential to detect pulsars hidden at the Galactic Centre and other extreme-scattering regions.

\subsubsection*{Acknowledgement}

Financial support by the European Research Council for the ERC Synergy Grant BlackHoleCam (Grant Agreement no. 610058) is gratefully acknowledged.

\end{document}